\documentclass[floats,prb,twocolumn,amsmath,superscriptaddress]{revtex4}

\usepackage{graphicx}
\usepackage{color}
\usepackage{dcolumn}

\begin{document}

\newcommand{\ket}    [1]{{|#1\rangle}}
\newcommand{\bra}    [1]{{\langle#1|}}
\newcommand{\braket} [2]{{\langle#1|#2\rangle}}
\newcommand{\bracket}[3]{{\langle#1|#2|#3\rangle}}
\newcommand{\red}{\textcolor{red}}
\newcommand{\blue}{\textcolor{blue}}
\newcommand{\green}{\textcolor{green}}
\newcommand{\cyan}{\textcolor{cyan}}
\newcommand{\magenta}{\textcolor{magenta}}

\def\pw{^{({\rm W})}}
\def\ph{^{({\rm H})}}
\def\k{{\bf k}}
\def\R{{\bf R}}
\def\b{{\bf b}}
\def\q{{\bf q}}
\def\o{{\cal O}}
\def\e{{\cal E}}
\def\v{{\rm v}}
\def\pw{^{({\rm W})}}
\def\ph{^{({\rm H})}}
\def\la{\langle\kern-2.0pt\langle}
\def\ra{\rangle\kern-2.0pt\rangle}
\def\vt{\vert\kern-1.0pt\vert}
\def\D{{D}\ph}
\def\n{{\cal N}}
\def\u{{\cal U}}

\title{
Density-functional investigation of the rhombohedral to simple cubic phase 
transition of arsenic}

\author{Patricia Silas}
\affiliation{Theory of Condensed Matter, Cavendish Laboratory,
University of Cambridge, JJ Thomson Avenue, Cambridge CB3 0HE, United Kingdom}

\author{Jonathan R. Yates}
\affiliation{Theory of Condensed Matter, Cavendish Laboratory,
University of Cambridge, JJ Thomson Avenue, Cambridge CB3 0HE, United Kingdom}

\author{Peter D. Haynes}
\affiliation{Departments of Materials and Physics,
Imperial College London, Exhibition Road, London SW7 2AZ, United Kingdom}

\date{\today}
\begin{abstract}
We report on our investigation of the crystal 
structure of arsenic under compression, focusing primarily on
the pressure-induced A7~$\to$~simple cubic (sc) phase transition. 
The two-atom rhombohedral unit cell is subjected to pressures
ranging from 0~GPa to 200~GPa; for each given pressure, 
cell lengths and angles, as well as atomic positions, 
are allowed to vary until the fully relaxed structure is obtained.
We find that the nearest and next-nearest neighbor distances
give the clearest indication of the occurrence of a structural
phase transition.
Calculations are performed using the local density approximation
(LDA) and the PBE and PW91 generalized gradient
approximations (GGA-PBE and GGA-PW91) for the exchange-correlation functional.
The A7~$\to$~sc
transition is found to occur at 21$\pm$1~GPa in the LDA,
at 28$\pm$1~GPa in the GGA-PBE and at 29$\pm$1~GPa in the GGA-PW91;
no volume discontinuity
is observed across the transition in any of the three cases.
We use \mbox{$k$-point} grids as dense 
as 66$\times$66$\times$66 to enable us to present reliably converged
results for the A7~$\to$~sc transition of arsenic.
\end{abstract}


\maketitle

\section{Introduction}
\label{sec:intro}

It is currently de rigueur to examine, both experimentally and 
theoretically, the structure of materials subjected to extremely high
pressures.\cite{mcmahon_06}  The \mbox{group--V} semi-metals, which include arsenic,
are of particular interest as they all exhibit a distinctive low-symmetry
structure when uncompressed.  With pressure, these elements
experience transitions into structures of higher symmetry, with
unusual intermediate phases --- incommensurate or {\it host-guest}
structures --- occurring along the way.

The computational study of the high-pressure behavior
of materials necessitates the ability to properly examine displacive
phase transitions from one structure to another.
Some structural phase transitions are straight-forward enough that
the transition pressure can be determined by looking at where the
enthalpy-pressure curves of the two structures cross, or by using
the technique of the ``common tangent'' on the energy-volume curves
of the two structures.  However, the transition pressure cannot be
determined in this way if the enthalpy-pressure (or energy-volume)
curves of the two structures merge, as in the case of the 
A7~$\to$~sc transition of arsenic (and for the same
transition in the other \mbox{group--V} semi-metals).
So if the phase transition between the two structures is a smooth
one --- such that the energy differences between the two structures
in the region of the transition are extremely small --- then how
should such a transition be studied?  High levels of convergence
of the calculation of the quantities of interest are crucial.

In the past, the A7~$\to$~sc transition pressure of
arsenic has been the subject of experimental as well as 
theoretical dispute.  Experimentally, there has been
a long-standing question as to whether results obtained
by Beister~\cite{beister_90} for this transition pressure 
are correct over those of Kikegawa and Iwasaki~\cite{kikegawa_87}
--- our results support the findings of the former. 
Existing theoretical studies of arsenic
yield a wide range of possible values for the transition
pressure.  We believe that a great part of the reason
for this spread of values is inadequate \mbox{$k$-point}
sampling of the Brillouin zone.  The Fermi surface of
arsenic is extremely complex: using values
reported by Lin and Falicov~\cite{linandfalicov_66} for the
cross-sectional area of a \textit{neck} (the finest
feature of the hole Fermi surface), we estimate
that to sample the Fermi surface of arsenic
at ambient pressures such that all of the features
can be resolved --- and are subsequently correctly weighted  --- a
grid of at least 140$\times$140$\times$140 \mbox{$k$-points} would be
required.  In practice, it is not necessary to use grids as dense
as this but one would not know this without proper
and thorough convergence testing with respect to
\mbox{$k$-point} grid size and amount of smearing used.
As no such convergence testing of the A7~$\to$~sc transition
of arsenic appears in the literature, results of
computational studies to date are unreliable and
any agreement with experiment has been fortuitous.
We undertake indepth studies of the convergence
properties of this transition,\cite{silas_08}
which enable us to proffer reliably converged results.

This investigation is not only timely, given the
current levels of interest in the field of high-pressure research,
but it is also especially relevant to the study of
pressure-induced insulator to metal or
semi-metal to metal phase transitions.  
In particular, we find that inspection of such transitions
demands careful convergence of the quantities of interest with
respect to \mbox{$k$-point} sampling and smearing, yet
such testing is not always performed.

Using density functional theory (DFT) to study the high-pressure
properties of materials furthermore necessitates an awareness of how different
approximations to the exchange-correlation functional perform
at such a task; to understand the differences encountered in the use
of these functionals is in itself a motivation for undertaking
theoretical investigations of materials at high pressures.\cite{mujica_03}
We thus compare the performance of the LDA with that of
two GGA functionals, having achieved accurate results for each.

The paper is organized as follows.  Section II contains the background
material that pertains to our study, including a description of
the A7 phase, a review of the structural transformation
that yields the sc phase when a sufficiently high external pressure is
applied, and a survey of the relevant findings published prior to this work.
Section III deals with the computational details of our 
calculations.  In Sec. IV, we present and discuss our results,
contrasting them against those that appear in the literature.
We begin by examining, using the nearest and next-nearest neighbor
distances, the overall picture of the pressure-dependence of the
phases of arsenic for the LDA and the GGA cases; the pressure-dependence
of the lattice parameters is also explored.  Next we focus on the
A7~$\to$~sc transition in particular, discussing it in more detail
using energy-volume and pressure-volume curves.  In Sec. V, we provide
a brief discussion and conclusion.

\section{The Structures of Arsenic}
\label{sec:structure}

\begin{figure}
\begin{center}
\includegraphics[width=\columnwidth]{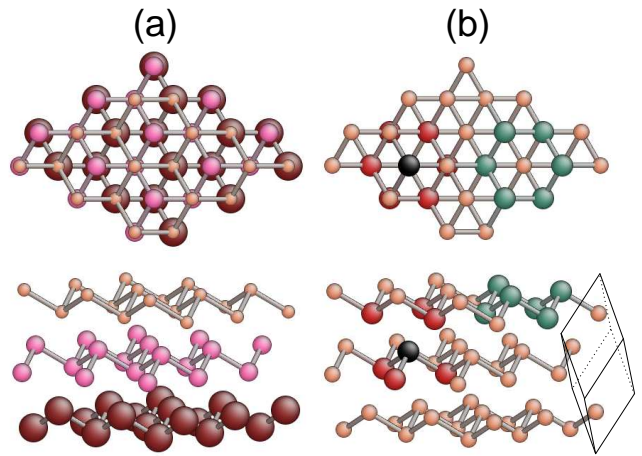}
\caption{(color online) The [111] direction is out of the page
in the top portion of the figure; in the bottom portion of
the figure it lies in the plane of the page and points upward.
\mbox{(a) Sizes} and colors of atoms depicted carry no significance, but are employed
merely to illustrate the ABC stacking of the A7 structure of arsenic. (b) The A7
structure consists of layers of 6-member buckled rings as displayed by the hexagon composed
of green arsenic atoms.  The buckled appearance of arsenic is due to the bonding of
nearest neighbors that occurs within each of the layers; the layers themselves are only
weakly bonded.  Nearest and next-nearest neighbor distances are illustrated using the
red atoms and are measured with respect to the black atom.  Nearest neighbor
distances exist between red and black atoms within the same layer; next-nearest
neighbor distances exist between red and black atoms from adjacent layers.  The two-atom
rhombohedral unit cell is also displayed. (This figure was created in part using
AtomEye.\cite{atomeye})
}
\label{fig: structure}
\end{center}
\end{figure}

At ambient pressures, arsenic is a covalently bonded compound existing
in the rhombohedral (A7) phase.
The A7 phase, belonging to space group~166, is a low-symmetry,
three-fold coordinated, layered structure (Fig.~\ref{fig: structure}).  
Nearest neighbors exist within a common layer;
bonding between nearest neighbors gives the structure its buckled appearance.
Adjacent layers, which contain next-nearest neighbors, are weakly bonded.
The stacking of the layers occurs in the [111] direction, as does the buckling.

The rhombohedral primitive cell of arsenic contains two atoms and is described by
the length of the primitive lattice vectors, $a$, 
the angle, $\alpha$, between each set of primitive lattice vectors, and by the 
atomic positional parameter, $z$, which determines the positions of the two
atoms along the cell's body diagonal.  The two atoms are located at (Wyckoff)
positions $(z,z,z)$ and $(-z,-z,-z)$ \cite{wyckoff} along the primitive lattice vectors,
where the atomic positional parameter $z$ is a fractional coordinate.
Together, $\alpha$ and $z$ characterize the degree of buckling experienced by
the A7 structure.

The A7 structure can be interpreted as a distortion of the six-fold coordinated
sc structure;\cite{falicovandgolin_65, needs_86, mattheiss_86}
a diagram charting the evolution of the sc structure into
the A7 structure can be found in Ref.~\onlinecite{beister_90}.
The sc lattice can be viewed as being
composed of two interpenetrating face-centered cubic (fcc) lattices.  The
vertices of a sc cell along the [111] direction contain one atom from
each of the fcc lattices.  A rhombohedral cell is superimposed onto this in such
a way that its origin lies central to, and contains, two of these alternate fcc
lattice atoms.  Its body diagonal is along the [111] direction
(the trigonal axis). Along this direction, the atomic positional parameter $z$
can be thought of as the ratio of the distance from the origin to the first atom,
divided by the length of the body diagonal.  A shear is then applied along the
trigonal axis so that the angle $\alpha$ between sets of primitive lattice
vectors decreases from its simple cubic value of $60^{\circ}$.
Finally, the two atoms are displaced from their original
positions and away from each other such that $z$ decreases from its sc
value of 0.25.  The periodicity of the atoms in the [111] direction is now doubled; pairs
of atoms in this direction repeat a pattern of being further apart and then nearer.
In this way, the coordination number is reduced from six to three, as is consistent
with the tendency of \mbox{group--V} semi-metals at normal pressures to form three bonds.

Subjected to sufficiently high pressures, arsenic undergoes a transition from 
semi-metallic A7 to metallic sc.
At lower pressures, the stability of the A7 structure against that of the sc structure
can be understood in terms of a Peierls-type distortion associated with the displacement
of the atoms along the [111] direction from their original positions within the simple
cubic lattice.  This displacement corresponds to the longitudinal acoustic phonon
mode at the point $R=\pi(1,1,1)/a$ on the boundary of the simple cubic Brillouin
zone.\cite{chang_86}  The breaking of symmetry causes a lowering of the energy of the
electron states near the Fermi level, with the result that a gap is opened up,
the band contribution to the total energy is reduced, and the structure
is stabilized against the distortion.
As the pressure is increased, ionic effects become more significant, and the Ewald
contribution to the total energy increases.  Changes in the Ewald energy eventually
dominate over changes in the band energy, the A7 structure is destabilized such
that the Ewald energy is lowered, and the sc structure is recovered.\cite{needs_86,
chang_86}
Within the framework of chemical bonding, at low pressures the nearly orthogonal
bonds of the A7 structure indicate weak s-p hybridization, where valence
electrons are localized within saturated covalent bonds.  The splitting between
s and p bands increases with pressure, until the s band is so deeply depressed
with respect to the p band that unsaturated orthogonal bonds form, resulting in
the metallic sc structure.\cite{mattheiss_86, littlewood_80}

It is sometimes convenient to use the nonprimitive hexagonal (trigonal) representation of
the unit cell of arsenic.  The trigonal cell 
is described by the lattice vectors ${\bf a}_{\,\text{hex}}$ and
${\bf c}_{\,\text{hex}}$, the trigonal axis which lies in
the [111] direction and coincides with the body diagonal of the rhombohedral
primitive cell, and by the same atomic positional parameter $z$.  The six atoms
contained within the cell are located at Wyckoff positions $\pm(0,0,z)rh$.\cite{wyckoff}
The lattice vector lengths $a_{\,\text{hex}}$ and $c_{\,\text{hex}}$ are related to the
length of the primitive lattice vector $a$
and angle $\alpha$ as follows:  $a_{\,\text{hex}}\!=\!2a\sin(\alpha/2)$, and
$c_{\,\text{hex}}\!=\!a\{3[1+2\cos(\alpha)]\}^{1/2}$.  Within this representation, $z$ can be thought
of as locating, as a fraction of ${\bf c}_{\,\text{hex}}$, the first atom from the origin along this axis.

From ambient conditions and with increasing pressure, the experimentally
observed transition sequence for the phases of 
arsenic is \mbox{A7~$\to$~sc~$\to$~{A}s-{III}~$\to$~bcc}, where bcc is the body-centered
cubic phase and \mbox{As-III} is an incommensurate structure.
In the literature, the values reported
for the pressure at which occurs the first structural transition of
arsenic, from semi-metallic rhombohedral to metallic simple cubic, are spread out
over a range of approximately 20~GPa.  It has been theoretically predicted that arsenic
should undergo a structural phase transition at 18~GPa;\cite{schirber_71} early
numerical studies of arsenic yielded for the transition pressure values of 35~GPa \cite{chang_86}
and 19~GPa.\cite{mattheiss_86} This last result corresponded to $V/V_{o}\simeq0.8$
(and was indeed found again very recently in a study of the lattice dynamics of arsenic
to be the volume ratio at which a transition occurs \cite{shang_07}),
where $V$ is the volume of the compressed cell
and $V_{o}$ is the uncompressed equilibrium volume.  Investigating the pressure-induced
superconductivity of arsenic, the experimental findings of Kawamura and Wittig
\cite{kawamura_85} suggest a transition
at 24~GPa, while those of Chen, ~\textit{et al.}~\cite{chen_92} support a
transition pressure of 32~GPa, though they obtain 36~GPa via the theoretical
component of their study.
Performing energy-dispersive and angle-dispersive powder X-ray diffraction studies of arsenic
up to 45~GPa, Kikegawa and Iwasaki \cite{kikegawa_87}
determined the A7~$\to$~sc transition to be 
of first order, occurring somewhere in the interval of \mbox{31.4--36.6~GPa} and resulting in a
cell volume discontinuity of $\Delta{V}/V_{T}\simeq5\%$,
where $\Delta{V}$ is the difference in the cell volume across transition,
and $V_{T}$ is the volume of the cell just prior to the transition. 
\mbox{Beister, \textit{et al.},
\cite{beister_90}} conducting an angle-dispersive powder X-ray diffraction study up to
33~GPa in addition to a Raman investigation, concluded that the transition occurs
at 25~GPa, and that to within their experimental uncertainty no
discernable change in volume is seen to occur:  $\Delta{V}/V_{T}<0.5\%$. 
More recently, theoretical investigations of the A7~$\to$~sc transition
of arsenic yielded for the transition pressure a range of $25\pm8$~GPa \cite{dasilva_97}
in one study, and 28~GPa \cite{haussermann_02} in another.
A molecular dynamics simulation of the A7~$\to$~sc transition
performed by Durandurdu \cite{durandurdu_05} yielded a transition pressure of 35~GPa
and a volume discontinuity of $3.2\%$, and most recently, transition pressures of
22~GPa \cite{feng_07} and 26.3~GPa \cite{zijlstra_08} have been found, with 
volume discontinuities of 0.8\% and 0.4\% respectively. 

Experiments aimed at investigating the higher-pressure phase transitions of arsenic were first
carried out by \mbox{Greene, \textit{et al.}.\cite{greene_95}}  Energy-dispersive powder
\mbox{X-ray} diffraction experiments on arsenic up to 122~GPa yielded
sc~$\to$~\mbox{As-III} and \mbox{As-III}~$\to$~bcc transition pressures
of 48$\pm$11~GPa and 97$\pm$14~GPa, respectively.  To within their experimental
error of 1$\%$, no measurable volume discontinuities for either phase transition
were observed.  
These results were reproduced rather closely in a theoretical study 
by H{\"{a}}ussermann,~\textit{et al.},\cite{haussermann_02} in which the two pressures were
found to be 43~GPa and 97~GPa. Reinterpration of diffraction patterns obtained in the
experiments of \mbox{Greene, \textit{et al.}} led Iwasaki \cite{iwasaki_commentongreene_97}
(see also Ref.~\onlinecite{iwasaki_97}) to suggest a tetragonal \mbox{Bi-III}-type 
or \mbox{Sb-II}-type structure, regarded as distorted bcc, for the
\mbox{As-III} phase.  More recently \mbox{McMahon, \textit{et al.},\cite{mcmahon_00}}
using angle-dispersive \mbox{X-ray} diffraction techniques,
determined both the \mbox{Bi-III} and \mbox{Sb-II} phases to be \mbox{Ba-IV}-type structures,
consisting of a body-centered tetragonal (bct) \textit{host}
structure containing a bct \textit{guest} component that is incommensurate with the 
host along the tetragonal $c$ axis.  They suggested that a modification of this structure
in which the host component is bct and the guest is monoclinic would be 
appropriate for \mbox{As-III}.  \mbox{Degtyareva, \textit{et al.} \cite{degtyareva_04}}
later found that the \mbox{As-III} structure closely resembles that of \mbox{Sb-IV},
and proposed that it is possible that arsenic may undergo an
incommensurate-to-incommensurate transition between the \mbox{As-III} and bcc phases
similar to the \mbox{Sb-IV}~$\to$~\mbox{Sb-II} transition observed in their study.

\section{Computational Details}

All calculations were performed using the CASTEP code.\cite{castep}  For the
exchange-correlation functional, we use the local density approximation (LDA)
\cite{ceperley_80} as parametrized by Perdew and Zunger, \cite{perdew_81} 
the Perdew-Burke-Ernzerhof generalized gradient approximation (abbreviated as GGA-PBE)\cite{perdew_96}
and the Perdew-Wang generalized gradient approximation (GGA-PW91).\cite{perdew_92}

The interaction between the core and valence states of arsenic is described
using scalar-relativistic ultra-soft pseudopotentials.\cite{vanderbilt_90} 
For the calculations to proceed efficiently, such pseudopotentials
require the use of two \mbox{FFT-grids}:
a \textit{standard} grid,
which is used to represent the pseudo wave-functions, and a \textit{fine} grid,
which is used to store the charge density, and thus the ultra-soft augmentation charge.
The size of the standard grid is determined by the \textit{cut-off energy} --- the
maximum energy accounted for in the plane-wave expansion of the wave-functions;
the fine grid must be chosen dense enough such that
the charge-density within
the atomic cores is reproduced with sufficient detail --- this
is a requirement to ensure
well-converged forces and stresses.
Convergence tests on our system at atmospheric pressure led us to
choose a cut-off energy of 450~eV; we chose our fine grid to be twice as dense as
our standard grid.  With these choices, our energies, forces and stresses are
converged to within 0.003~eV/atom, 0.001~eV/\AA\, and 0.01~GPa, respectively.

We apply pressures between 0~GPa and 200~GPa to the two-atom
rhombohedral unit cell of arsenic,
allowing cell lengths and angles in addition
to atomic positions to vary. 
For each applied pressure, the enthalpy is
minimized such that the system
is relaxed from the structure that exists at atmospheric 
pressure as determined experimentally by Schiferl.\cite{schiferl_69} The geometry
optimizations are made to continue until the differences in the forces and stresses
between iterations are less than 0.005~eV/\AA\, and 0.05~GPa, respectively. 
Proper convergence studies in electronic structure calculations are
important and the lack of such testing can lead
to severe inaccuracies.\cite{mehl_00}
Accordingly, we have thoroughly investigated the structural convergence
of our system in the region of the A7~$\to$~sc phase transition;
we have examined the convergence with respect to \mbox{$k$-point} grid size
and amount of cold-smearing \cite{marzari_99} used ---  the results
of this convergence testing can be found in the supplementary material provided.\cite{silas_08}
In the vicinity of the A7~$\to$~sc transition, the
\mbox{$k$-point} grids that have been studied are $n{\times}n{\times}n$, where $n$ is 24, 25, 26, 33,
50 and 66 (all \mbox{$k$-point} grids are unshifted, so that the gamma point
has been included in the calculations), using cold-smearings of 0.1, 0.2 and
0.5~eV in the LDA and 0.1~eV in the GGA-PBE.
At pressures outside the region of this phase transition (and
for all pressures studied in the case of the GGA-PW91), a \mbox{$k$-point} grid of
33$\times$33$\times$33 is used, along with a cold-smearing of 0.1~eV.
The symmetry of the initial A7 structure is maintained throughout the relaxation,
as a highly symmetric path is assumed for the A7~$\to$~sc transition, with
atoms moving only along the [111] direction.  Although it is irrelevant whether
this assumption is valid for the \mbox{As-III} phase, because no attempt has been made
here to properly model this incommensurate structure, it seems reasonable that
imposing these constraints on the symmetry of the system will not affect the
transition pressures observed since the incommensurate phase is bounded by two phases
of higher symmetry than the original A7 structure.

Finally, all calculations are performed at $T=0$\,K, and
the zero-point vibrational energy, as well as the spin-orbit interaction,
are neglected.

\section{Results and Discussion}
\subsection{Overview of the transitions of arsenic}

\begin{figure*}
\begin{center}
\includegraphics[width=\textwidth]{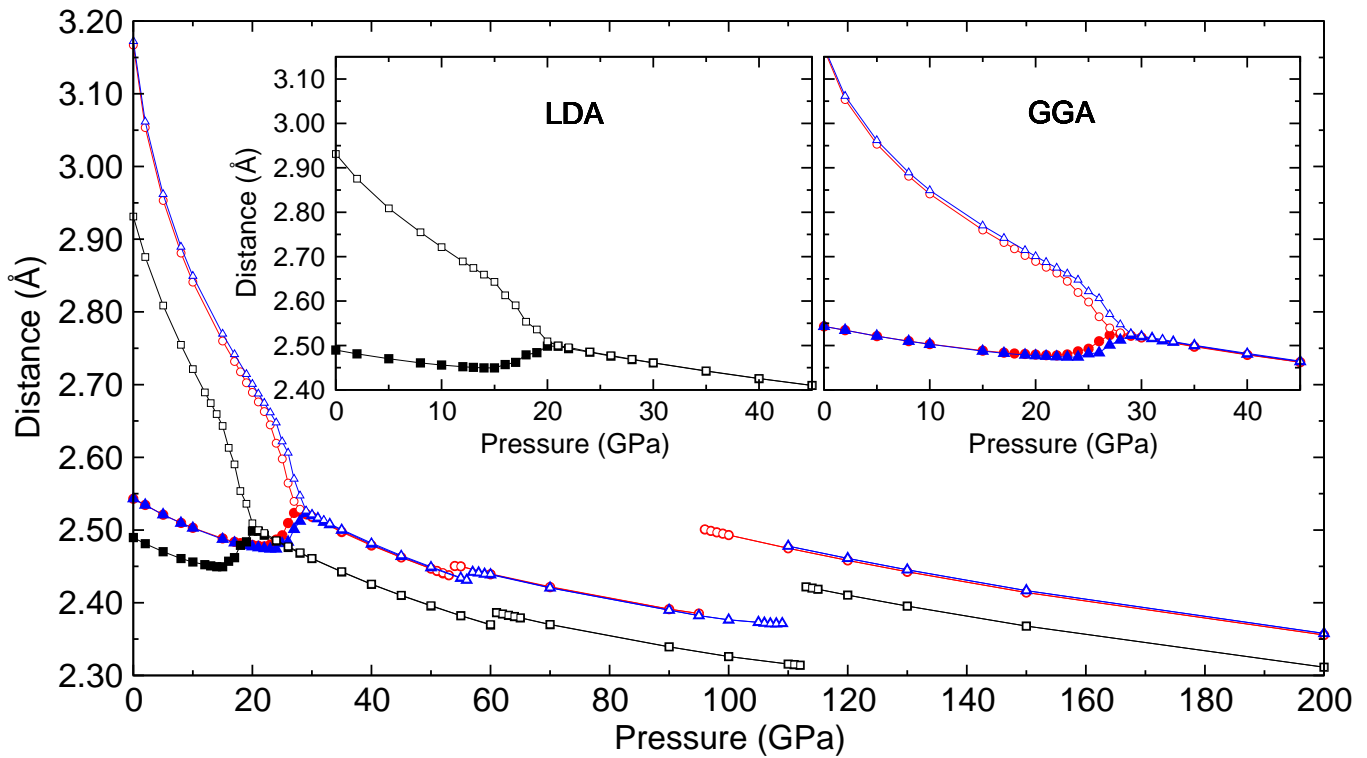}
\caption{(color online) Nearest neighbor (solid symbols) and next-nearest neighbor
(open symbols) distances as a function of pressure for the LDA (black squares),
for the GGA-PBE (red circles) and for the GGA-PW91 (blue triangles).
The A7~$\to$~sc, sc~$\to$~\textit{incomm.}
and \textit{incomm.}~$\to$~bcc transitions are evident, where the insets
emphasize the A7~$\to$~sc transition in both the LDA and the GGA.
(Calculations performed using a 33$\times$33$\times$33 \mbox{$k$-point}
grid and a cold-smearing of 0.1~eV.)}
\label{fig: nnn_and_nn}
\end{center}
\end{figure*}

We first take an overview of the behavior of the nearest and
next-nearest neighbor distances as the pressure is increased from 0~GPa to
200~GPa.  It has already been stated that for the
A7~$\to$~sc phase transition in particular, the behavior of these two
quantities gives the clearest indication of when a structural phase transition has
occurred.  Fig.~\ref{fig: nnn_and_nn} reveals the results we obtained for 
each of the exchange-correlation functionals studied, and includes separate close-ups of the
A7~$\to$~sc phase transition for the LDA and for the GGA.  Indeed, it is quite clear when this 
transition occurs in all cases, as it happens when the nearest and next-nearest neighbor
distances become equal.  From the insets of Fig.~\ref{fig: nnn_and_nn}, we can
conclude that the A7~$\to$~sc phase transition of arsenic occurs at 21$\pm$1~GPa
in the LDA, at 28$\pm$1~GPa in the GGA-PBE and at 29$\pm$1~GPa in the GGA-PW91
(note that the difference in the transition pressure that arises from the use of the GGA-PW91
over the GGA-PBE is negligible as it is within the uncertainty of 1~GPa).
This result is more consistent with the experimental findings
of \mbox{Beister, \textit{et al.},\cite{beister_90}} who found the phase
transition to occur at 25$\pm$1~GPa,
than it is with those of Kikegawa and Iwasaki,\cite{kikegawa_87}
who found it to occur in the range of \mbox{31.4--36.6}~GPa;
it furthermore disagrees with the outcome of simulations performed recently
by Durandurdu.\cite{durandurdu_05} 

From this same figure we note incidentally that the GGA nearest and next-nearest neighbor distances
are always larger than those resulting from the LDA.  This is due to the well-known
fact that the LDA tends to overbind systems, whereas the GGA tends
to underbind them.\cite{seifert_95}  Thus cell volumes, cell parameters, and
distances contained within the cell will tend to be lower for the LDA
than for the GGA. 
 
Our unit cell contains two atoms and is periodically repeated.  
We cannot hope to say anything about the incommensurate
phase that occurs in the pressure-induced progression of phases
of arsenic, and it is not our intention to do so in this paper;
to acknowledge this point, we will use the abbreviation \textit{incomm.}
to represent what we find between the sc and bcc phases, and subsequently write
the progression of phases as \mbox{A7~$\to$~sc~$\to$~\textit{incomm.}~$\to$~bcc}.
We mention as an aside that we have chosen to work with the
two-atom unit cell of arsenic
in the interest of undertaking efficient calculations,
but as we are guided by experiment and not primarily
focused in this study on the incommensurate structure,
a two-atom unit cell is sufficient for our needs.
It is possible that were we to use unit cells
containing arbitrary numbers of atoms we might
discover other more complex or more closely-packed
structures of arsenic at pressures
higher than have yet been attained experimentally ---
this is an interesting point and although outside the
scope of this work, would be an excellent candidate
for the application of fully randomized structure
prediction.\cite{pickard_06}

Having stated that we do not expect to reproduce the incommensurate
phase of arsenic, we can still use our system to approximate the pressure at
which the system ceases to be sc, as well as to approximate 
the pressure at which it becomes bcc, and 
compare with the most recent experimental findings.
Thus, referring again to Fig.~\ref{fig: nnn_and_nn}, we see that
in the LDA the sc~$\to$~\textit{incomm.}
phase transition appears to occur at \mbox{60--61}~GPa and the \textit{incomm.}~$\to$~bcc
at \mbox{112--113}~GPa.  In the GGA-PBE it is \mbox{53--54}~GPa and \mbox{95--96}~GPa,
and in the GGA-PW91 it is \mbox{56--57}~GPa and \mbox{109--110}~GPa.
Although our findings for the sc~$\to$~\textit{incomm.} transition
are about 10~GPa higher than those of Ref.~\onlinecite{haussermann_02},
and fall into the high-end of the range of values given in
Ref.~\onlinecite{greene_95},  our results for the \textit{incomm.}~$\to$~bcc
transition agree closely with those from both works.  Along with
the findings that are present in the literature, our results for the transition
pressures of arsenic are summarized in Table~\ref{table: pressures}.

\begin{table}
\caption{
\label{table: pressures}
Transition pressure $P_{T}$, reduced transition volume $V_{T}/V_{o}$, where $V_{T}$ is 
the volume just prior to the transition and $V_{o}$ is the uncompressed equilibrium volume,
and percentage volume change across the A7~$\to$~sc transition. Below are listed transition pressures
for the higher pressure transitions of arsenic. Parametrizations of the 
generalized gradient approximation to the exchange-correlation functional
that appear below are PBE~\cite{perdew_96} and PW91.\cite{perdew_92} (Abbreviations:  PP -- pseudopotential approach, LAPW -- linear augmented plane wave approach.)} 
\begin{ruledtabular}
\begin{tabular}{lccc}
 $A7~\to~sc$ & $P_{T}~\mathrm{(GPa)}$ & $V_{T}/V_{o}$ & $\Delta{V}/V_{T}~(\%)$ \\
\hline
Theory: \\
This work, LDA & $21\pm1$ \\
This work, PBE & $28\pm1$ \\
This work, PW91 & $29\pm1$ \\
Ref.~\onlinecite{chang_86} (PP, LDA) & 35 & 0.72  \\
Ref.~\onlinecite{mattheiss_86} (LAPW, LDA) & 19 & 0.8  \\ 
Ref.~\onlinecite{chen_92} (PP, LDA) & 36  \\
Ref.~\onlinecite{dasilva_97} (PP, LDA) & $25\pm8$ & 0.78 & 1 \\
Ref.~\onlinecite{haussermann_02} (PP, PW91) & 28 &  & 0.8 \\
Ref.~\onlinecite{durandurdu_05} (PP, LDA) & 35 &  & 3.2 \\
Ref.~\onlinecite{feng_07} (PP, PBE) & 22 &  & 0.8 \\
Ref.~\onlinecite{shang_07} (PP, PW91) &  & 0.8  \\
Ref.~\onlinecite{zijlstra_08} (LAPW, PBE) & 26.3 &  & 0.4 \\
Experiment: \\
Ref.~\onlinecite{schirber_71} & 18  \\
Ref.~\onlinecite{kawamura_85} & 24 \\
Ref.~\onlinecite{kikegawa_87} & 31.4 -- 36.6 & 0.744 & 5 \\
Ref.~\onlinecite{beister_90} & $25\pm1$ & 0.772 & 0.5 \\
Ref.~\onlinecite{chen_92} & 32 \\
\end{tabular}
\begin{tabular}{lcc}
 $P_{T}~\mathrm{(GPa)}$ & $sc~\to~\textit{incomm.}$ & $\textit{incomm.}~\to~bcc$ \\
\hline
Theory: \\
This work, LDA & $60.5\pm0.5$ & $112.5\pm0.5$ \\
This work, PBE & $53.5\pm0.5$ & $95.5\pm0.5$ \\
This work, PW91 & $56.5\pm0.5$ & $109.5\pm0.5$ \\
Ref.~\onlinecite{haussermann_02} (PP, PW91) & 43 & 97 \\
Experiment: \\
Ref.~\onlinecite{greene_95} & $48\pm11$ & $97\pm14$ \\
\end{tabular}
\end{ruledtabular}
\end{table}

\begin{figure*}
\begin{center}
\includegraphics[width=\textwidth]{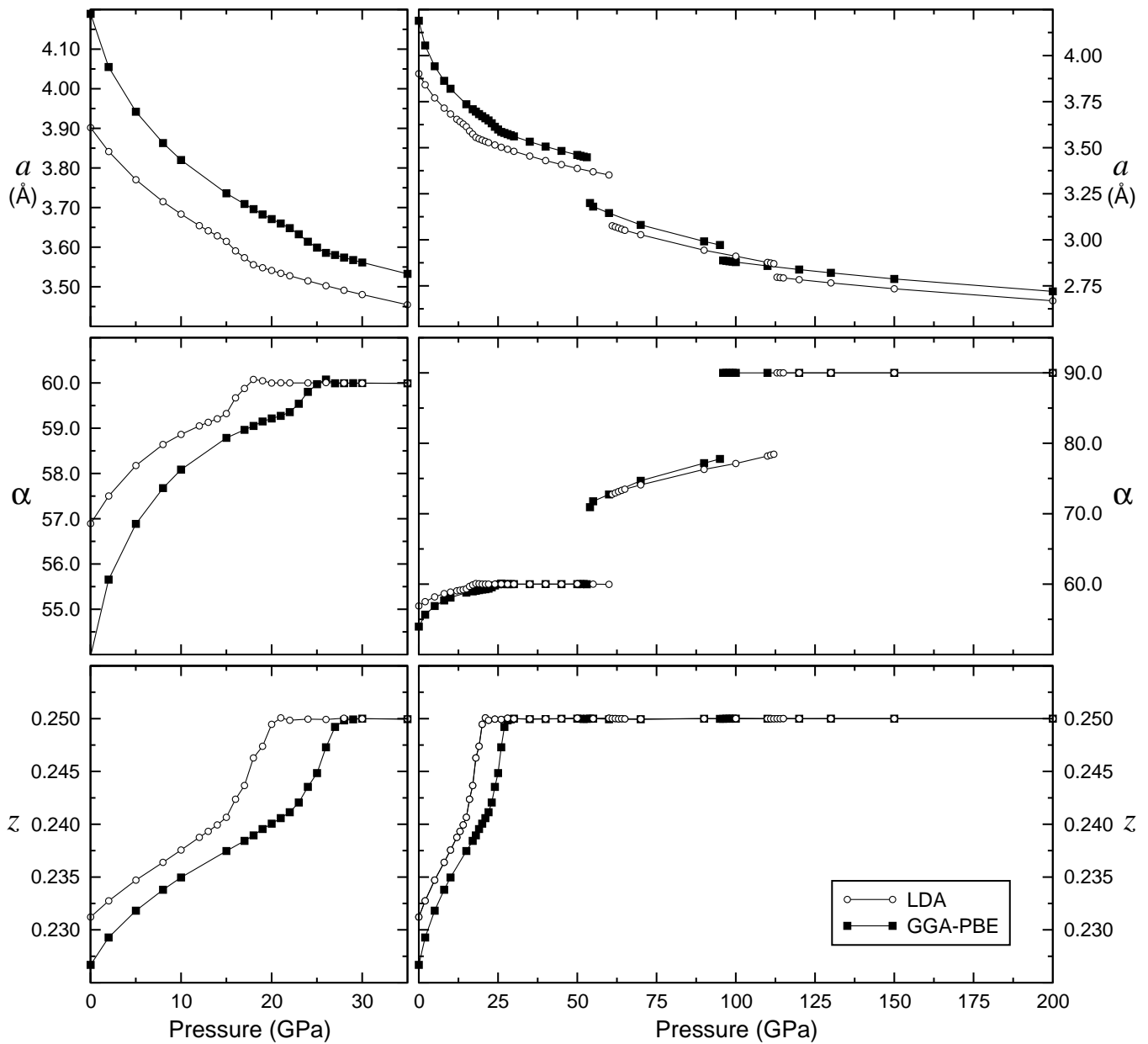}
\caption{Lattice parameter $a$ (top tier), cell angle $\alpha$ (middle tier) and
atomic positional parameter $z$ (bottom tier) as a function of pressure for the 
LDA (open circles) and for the GGA-PBE (filled squares). The left-hand panels
emphasize the {A}7~$\to$~sc transition, while in addition the sc~$\to$~\textit{incomm.}
and \textit{incomm.}~$\to$~bcc transitions are apparent in the top two tiers of the
right-hand panels. (Calculations performed using a 33$\times$33$\times$33 \mbox{$k$-point}
grid and a cold-smearing of 0.1~eV.)}
\label{fig: a_alpha_z}
\end{center}
\end{figure*}

We break down the nearest and next-nearest neighbor distances into their
constituent components in Fig.~\ref{fig: a_alpha_z}, which illustrates the 
behavior of the lattice parameter $a$ (top tier), the cell angle $\alpha$
(middle tier), and the atomic positional parameter $z$ (bottom tier), 
for the LDA and the GGA-PBE.
Each tier of Fig.~\ref{fig: a_alpha_z} is composed of a close-up
of the A7~$\to$~sc phase transition (on the left) along with the behavior
of the quantity over the entire range of pressures studied (on the right).
Respectively, the high symmetry values of $\alpha$ and $z$, as can be confirmed in this
figure, are 60$^\circ$ and 0.25 in the simple cubic case, and 90$^\circ$ and 0.25
in the bcc case.
The sc phase is of higher symmetry
than the A7 phase, and because we have constrained the symmetry
of the system, once the sc phase has been reached the
atomic positional parameter $z$ remains equal to 0.25.  Thus, after the
A7~$\to$~sc phase transition has occurred, it is the angle $\alpha$
that determines the phase of the structure as the applied pressure is increased.

The first point to observe upon inspection of Fig.~\ref{fig: a_alpha_z}
is that the A7~$\to$~sc transition
appears to be continuous with respect to all three quantities: $a$ (top), $\alpha$
(middle) and $z$ (bottom).  
We see from the set of right hand panels in the figure
that the sc~$\to$~\textit{incomm.} and the \textit{incomm.}~$\to$~bcc transitions
appear to be discontinuous with respect to lattice parameter $a$ and
cell angle $\alpha$, for both the LDA and the GGA; in particular, these discontinuities
subsequently manifest themselves in the nearest and next-nearest neighbor distances
at these two transitions as can be seen in Fig.~\ref{fig: nnn_and_nn}, and indicate
when the cell angle has jumped away from 60$^\circ$ in the case
of the former transition, and to 90$^\circ$ in the case of the latter.
Before discussing the behavior with pressure of the cell parameters further,
we first inspect more closely the A7~$\to$~sc transition.

\subsection{The A7~$\to$~sc transition of arsenic}

Examining the left hand panels of Fig.~\ref{fig: a_alpha_z}, we see
that $\alpha$ and $z$ reach their high-symmetry values
at lower pressures in the case of the LDA than in the GGA; this is to
be expected as we noted earlier that the LDA A7~$\to$~sc 
transition pressure is lower than the GGA transition pressure.
However, we also note that in both cases, \mbox{$\alpha$ reaches 60$^\circ$} at a slightly
lower pressure than \mbox{$z$ reaches 0.25}. This result was also found by \mbox{Seifert,
\textit{et al.}\cite{seifert_95}} in their study of the A7~$\to$~sc transition
of antimony.  In fact, there is no reason to assume that $\alpha$ and $z$
must reach their high-symmetry values at the same pressure --- yet
we cannot rule out the possibility that this pressure discrepancy 
may be an artifact resulting from the use of smearing
in our calculations.  The difference in pressures between
which $\alpha$ and $z$ reach their high-symmetry values is slightly more
pronounced for the LDA (approximately 3~GPa) than it is for the GGA
(approximately 2~GPa for both the GGA-PBE and the GGA-PW91).
The curves of the \mbox{lattice parameter $a$} and 
the \mbox{cell angle $\alpha$} seemingly progress together.
Thus the cell reaches its high-symmetry shape ``before'' the transition
to sc takes place; since both \mbox{$\alpha$ and $z$} must reach their high-symmetry
values for the transition to occur, the A7~$\to$~sc transition pressure is 
consequently determined by the \mbox{atomic positional parameter $z$}.  Thus we
see that $z$ reaches 0.25 at approximately 21~GPa in the LDA and at 
approximately 28~GPa in the GGA-PBE.

\begin{figure*}
\begin{center}
\includegraphics[width=\textwidth]{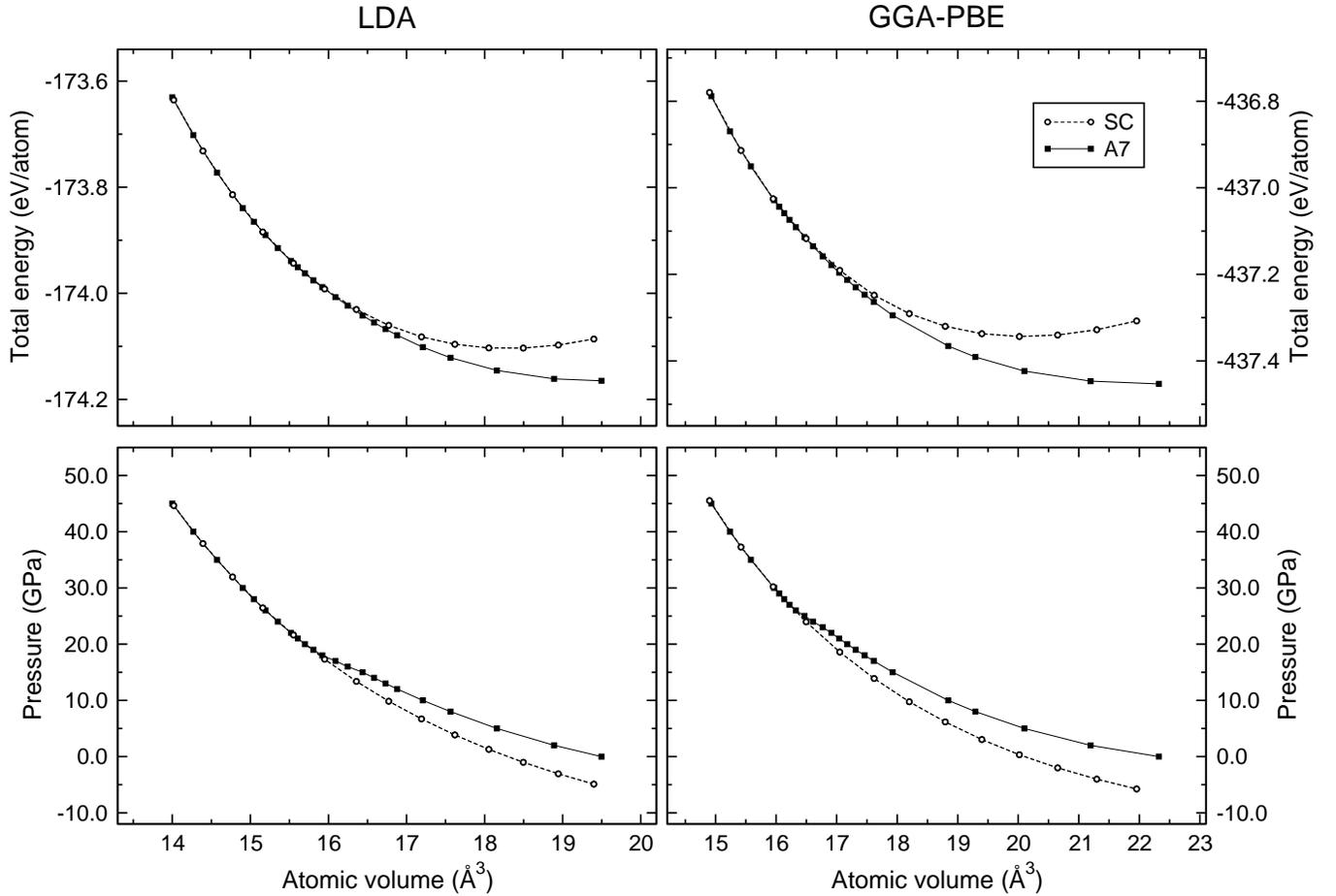}
\caption{Energy-volume (top) and pressure-volume (bottom) curves of the
A7 phase (filled squares) and of the simple cubic phase 
(open circles) for the LDA (left) and for the GGA-PBE (right).
For each of the four plots, the A7 and simple cubic curves merge,
and there are no discontinuities present suggesting that the A7~$\to$~sc
transition of arsenic is of second order. (Calculations performed using
a 33$\times$33$\times$33 \mbox{$k$-point} grid and a cold-smearing of 0.1~eV.)}
\label{fig: EvsV_PvsV}
\end{center}
\end{figure*}

We consider next, for the LDA and the GGA-PBE,
the energy-volume and pressure-volume curves of the A7
and sc structures over a pressure range of approximately \mbox{0--45}~GPa,
which includes only the A7~$\to$~sc phase transition
(Fig.~\ref{fig: EvsV_PvsV}).
In all cases, the A7 and sc curves merge once
the A7~$\to$~sc phase transition has occurred, in other words,
once the volume of the cell has been compressed beyond a particular value.
This contradicts the recent results of Durandurdu.\cite{durandurdu_05}
However, the sc~$\to$~A7 transformation is the result of two
continuous distortions, so we would expect this transition
also to be smooth and continuous.  We do not expect to see metastable
states, and we observe none --- this agrees with the experimental
findings both of \mbox{Beister, \textit{et al.},\cite{beister_90}} and 
of Kikegawa and Iwasaki,\cite{kikegawa_87} yet it disagrees with
the theoretical findings of \mbox{Da Silva, \textit{et al.}\cite{dasilva_97}}
We observe no discontinuities in these curves and so our results suggest that
the A7~$\to$~sc phase transition is of second order.  Experimentally, both
\mbox{Beister, \textit{et al.}\cite{beister_90}} and
Kikegawa and Iwasaki~\cite{kikegawa_87} report that the A7~$\to$~sc transition
is of first order, where to within an experimental error of 0.5\%
no volume discontinuity is detected in the former study, whereas
a volume discontinuity of 5\% across the transition is found in the latter study.
\mbox{Da Silva, \textit{et al.}\cite{dasilva_97}} report a transition that is
weakly first order, and a volume discontinuity of less than 1\%. 
We believe that if this transition appeared in the past to be of first order
it was because of the coarse grained nature of the investigations performed
at the time.

On the matter of the discrepancies that exist between our study
and, for example, those of Durandurdu~\cite{durandurdu_05} and
\mbox{Da Silva, \textit{et al.},\cite{dasilva_97}} it is clear
that in those investigations the \mbox{$k$-point} sampling
of the Brillouin zone was insufficient:
the most dense grid employed by \mbox{Da Silva, \textit{et. al.}\cite{dasilva_97}}
in their study of arsenic was 13$\times$13$\times$13 ---
Durandurdu~\cite{durandurdu_05} uses only the gamma point for a unit cell containing
250 atoms, roughly corresponding to using a 5$\times$5$\times$5 \mbox{$k$-point} grid for
for a two-atom unit cell.  Our investigations reveal that these calculations could not
have been converged.  Moreover, in addition to ensuring that the Brillouin zone is properly sampled, we
believe that choosing an appropriate amount of smearing is also very important (once again
consult the supplementary material for an indepth
study of the convergence properties with respect to
\mbox{$k$-point} sampling and smearing of the A7~$\to$~sc
transition of arsenic).\cite{silas_08}

Another point to be made about the insets of Fig.~\ref{fig: nnn_and_nn},
is that at pressures just below the A7~$\to$~sc
transition, the nearest neighbor distance actually increases with pressure.
This occurs over the range of approximately \mbox{15--20}~GPa in the
LDA and approximately \mbox{22--27}~GPa in the GGA-PBE.
The onset of this feature coincides with the point at which 
the $a$, $\alpha$ and $z$ versus pressure curves all begin to 
experience a kink or slight change of direction (see left-hand panels
of Fig.~\ref{fig: a_alpha_z}).
These pressure intervals also correspond to the regions where
the A7 pressure-volume curves of Fig.~\ref{fig: EvsV_PvsV} 
start dipping down (from right to left) toward those of the simple cubic phase.
As the volume does not increase
discontinuously across the transition, there must be significant atomic
restructuring going on for the nearest neighbor distance 
to increase as the transition is approached. 
Although their nearest
and next-nearest neighbor distances were estimates calculated according
to an empirical relation for the atomic positional parameter,
\mbox{Beister, \textit{et al.} \cite{beister_90}} noticed this point
as well and noted that such behavior is not unexpected, as there
is a tendency in covalently bonded solids to display increased 
nearest neighbor distances when undergoing pressure-induced
breaking of directional bonds, accompanied by increased coordination
numbers.

We have performed fits of the energy-volume curves using both the third-order
Birch-Murnaghan~\cite{birch_47} and Vinet~\cite{vinet_87} equations to yield values
for the bulk modulus, $B_{o}$, pressure derivative of the bulk
modulus, $B_{o}'$, and equilibrium volume, $V_{o}$, of arsenic
for both the A7 and sc structures.  We find that for the A7 phase,
the values we obtain for $B_{o}$ and $B_{o}'$ are extremely
sensitive to the number of points included in the fit; this was also
found to be the case for Needs,~\textit{et al.}\cite{needs_86} 
It is also somewhat the case for the sc portion of the A7 curve, so we have
quoted the values that we obtain when we
fit curves resulting from compressions of the exact sc structure,
rather than from geometry optimizations.
Table~\ref{table: fits} summarizes our findings, 
including the results we obtained for the equilibrium cell parameters
from our geometry optimizations of arsenic at zero pressure,
and compares them to those published in previous studies.

\begin{table*}
\caption{
\label{table: fits}
Lattice parameters for the uncompressed arsenic structure (A7), in addition to the equilibrium
volume ($V_{o}$), bulk modulus ($B_{o}$) and pressure derivative of the bulk modulus ($B_{o}'$). 
Those for the uncompressed simple cubic structure of arsenic are listed further below.
Parametrizations of the
generalized gradient approximation to the exchange-correlation functional
that appear below are PBE,\cite{perdew_96} PW86\cite{perdewandwang_86}
and PB.\cite{perdew_86,becke_88}
(Abbreviations: PP -- pseudopotential approach, LAPW -- linear augmented plane wave approach,
BM -- fitted with the third order Birch-Murnaghan\cite{birch_47} equation, V -- fitted
with the Vinet\cite{vinet_87} equation.)}
\begin{ruledtabular}
\begin{tabular}{lcccccc}
 A7 structure & $a$~{(\AA)} & $\alpha~(^\circ$) & $z$ & $V_{o}$~(\AA$^{3}$) & $B_{o}$~(GPa) & $B_{o}'$ \\
\hline
Theory: \\
This work, LDA & 3.902 & 56.89 & 0.231 & $19.48(1)^{\mathrm{BM,V}}$ & $60.9(1.6)^{\mathrm{BM,V}}$ & $4.6(4)^{\mathrm{BM}}$, $4.8(4)^{\mathrm{V}}$ \\ 
This work, PBE & 4.189 & 53.96 & 0.227 & $22.22(5)^{\mathrm{BM,V}}$ & $36.9(1.6)^{\mathrm{BM,V}}$ & $6.1(7)^{\mathrm{BM}}$, $6.25(55)^{\mathrm{V}}$ \\
This work, PW91 & 4.195 & 53.90 & 0.227 & $22.30(2)^{\mathrm{BM,V}}$ & $36.9(1.6)^{\mathrm{BM,V}}$ & $6.2(5)^{\mathrm{BM}}$, $6.3(4)^{\mathrm{V}}$ \\
Ref.~\onlinecite{needs_86} (PP, LDA) & 4.017 & 56.28 & 0.230 & 20.95 & 43  \\
Ref.~\onlinecite{needs_87} (PP, LDA) &  & & 0.24  \\
Ref.~\onlinecite{mattheiss_86} (LAPW, LDA) & 4.084 & 55.9 & 0.2294 & 21.8 & 77  \\
Ref.~\onlinecite{seifert_95} (PP, LDA) & 4.0027 & 56.24 & 0.230 & 20.70 & 52  \\
\hspace{0.44 in}(PP, PB) & 4.32035 & 52.44 & 0.225 & 23.45 & 36 \\
\hspace{0.44 in}(PP, PW86) & 4.57449 & 52.01 & 0.224 & 27.49  \\
Ref.~\onlinecite{dasilva_97} (PP, LDA) & 4.031 & 56.3 & 0.230 & 21.18 & 59(1) & 4.2(3) \\
Ref.~\onlinecite{durandurdu_05} (PP, LDA) & & & & 23.57 & 64.49 & 3.99 \\
Ref.~\onlinecite{feng_07} (PP, PBE) & 4.182 & 53.13 & 0.225 & 21.70 & 38.4 & 4.34 \\
Ref.~\onlinecite{zijlstra_08} (LAPW, PBE) & 4.246 & 53.31 & 0.226 & 22.82 \\
Experiment: \\
Ref.~\onlinecite{schiferl_69} & 4.1018(1) & 55.554(1) & 0.22764(4) & 21.303 \\
Ref.~\onlinecite{morosin_72} & & & 0.22728(7) \\
Ref.~\onlinecite{kikegawa_87} & 4.133 & 54.12 & 0.227 & 21.53 & 55.6 & 4.4 \\
Ref.~\onlinecite{beister_90} & & 54.13 & & & 58(4) & 3.3(4) \\
Ref.~\onlinecite{greene_95} & & & & & 56(3) & 3.7(2) \\
\end{tabular}
\begin{tabular}{lcccc}
 Simple cubic structure & $a_{\mathrm{cub}}$~(\AA) & $V_{o}$~(\AA$^{3}$) & $B_{o}$~(GPa) & $B_{o}'$ \\
\hline
Theory: \\
This work, LDA & 2.635 & $18.30^{\mathrm{BM,V}}$ & $94.53(16)^{\mathrm{BM}}$, $93.57(32)^{\mathrm{V}}$
 & $4.29(2)^{\mathrm{BM}}$, $4.48(3)^{\mathrm{V}}$  \\
This work, PBE & 2.719 & $20.10^{\mathrm{BM,V}}$ & $78.99(3)^{\mathrm{BM}}$, $78.51(16)^{\mathrm{V}}$
 & $4.318(4)^{\mathrm{BM}}$, $4.55(2)^{\mathrm{V}}$  \\
This work, PW91 & 2.722 & $20.17^{\mathrm{BM,V}}$ & $78.84(4)^{\mathrm{BM}}$, $78.32(19)^{\mathrm{V}}$
 & $4.297(3)^{\mathrm{BM}}$, $4.53(2)^{\mathrm{V}}$  \\
Ref.~\onlinecite{needs_86} (PP, LDA) & 2.687 & 19.4 & 104 & 4.4 \\
Ref.~\onlinecite{needs_87} (PP, LDA) & 2.68 & 19.25 & 122 & 2.32  \\
Ref.~\onlinecite{mattheiss_86} (LAPW, LDA) & 2.717 & 20.06 & 87  \\
Ref.~\onlinecite{dasilva_97} (PP, LDA) & 2.704 & 19.77 & 92(3) & 4.2(3) \\
Ref.~\onlinecite{durandurdu_05} (PP, LDA) & 2.822 & 22.47 & 74.15 & 3.2 \\
Ref.~\onlinecite{feng_07} (PP, PBE) &  2.674 & 19.12 & 88.2 & 3.93 \\
Ref.~\onlinecite{zijlstra_08} (LAPW, PBE) & 2.731 & 20.37 & 78.107 & 4.317 \\
\end{tabular}
\end{ruledtabular}
\end{table*}

\section{Conclusions}
\label{sec:conclusion}

We have performed simulations of the two-atom unit
cell of arsenic under compression to investigate
its high-pressure behavior and compare with experiment.
In the matter of the A7~$\to$~sc transition,
our results strongly support the experimental findings of
Beister~\cite{beister_90} over those
of Kikegawa and Iwasaki;~\cite{kikegawa_87}
we furthermore find this 
transition to be of second order.

Our main critique of the current literature is that we believe
that any results published until now concerning this semi-metal to metal
phase transition are unconverged (as no thorough convergence
studies have ever been published), and that any agreement
with experiment has merely been fortuitous.
This study has enabled us to present reliably converged results for the A7~$\to$~sc
transition of arsenic. 

Using DFT to study any pressure-induced structural phase transition involving a
metal demands convergence testing similar to that which we have carried out;~\cite{silas_08}
for such purposes,
\mbox{$k$-point} sampling and smearing must be considered thoroughly.

\section{Acknowledgments}
We thank Richard Needs for helpful discussions.  Computing resources
were provided by the Cambridge High Performance Computing Service (HPCS).
 
\bibliographystyle{apsrev}

\begin{thebibliography}{45}
\expandafter\ifx\csname natexlab\endcsname\relax\def\natexlab#1{#1}\fi
\expandafter\ifx\csname bibnamefont\endcsname\relax
  \def\bibnamefont#1{#1}\fi
\expandafter\ifx\csname bibfnamefont\endcsname\relax
  \def\bibfnamefont#1{#1}\fi
\expandafter\ifx\csname citenamefont\endcsname\relax
  \def\citenamefont#1{#1}\fi
\expandafter\ifx\csname url\endcsname\relax
  \def\url#1{\texttt{#1}}\fi
\expandafter\ifx\csname urlprefix\endcsname\relax\def\urlprefix{URL }\fi
\providecommand{\bibinfo}[2]{#2}
\providecommand{\eprint}[2][]{\url{#2}}

\bibitem[{\citenamefont{Mc{M}ahon and Nelmes}(2006)}]{mcmahon_06}
\bibinfo{author}{\bibfnamefont{M.~I.} \bibnamefont{Mc{M}ahon}}
  \bibnamefont{and} \bibinfo{author}{\bibfnamefont{R.~J.}
  \bibnamefont{Nelmes}}, \bibinfo{journal}{Chem. Soc. Rev.}
  \textbf{\bibinfo{volume}{35}}, \bibinfo{pages}{943} (\bibinfo{year}{2006}).

\bibitem[{\citenamefont{Beister et~al.}(1990)\citenamefont{Beister,
  Str{\"{o}}ssner, and Syassen}}]{beister_90}
\bibinfo{author}{\bibfnamefont{H.~J.} \bibnamefont{Beister}},
  \bibinfo{author}{\bibfnamefont{K.}~\bibnamefont{Str{\"{o}}ssner}},
  \bibnamefont{and} \bibinfo{author}{\bibfnamefont{K.}~\bibnamefont{Syassen}},
  \bibinfo{journal}{Phys. Rev. B} \textbf{\bibinfo{volume}{41}},
  \bibinfo{pages}{5535} (\bibinfo{year}{1990}).

\bibitem[{\citenamefont{Kikegawa and Iwasaki}(1987)}]{kikegawa_87}
\bibinfo{author}{\bibfnamefont{T.}~\bibnamefont{Kikegawa}} \bibnamefont{and}
  \bibinfo{author}{\bibfnamefont{H.}~\bibnamefont{Iwasaki}},
  \bibinfo{journal}{J. Phys. Soc. Jpn.} \textbf{\bibinfo{volume}{56}},
  \bibinfo{pages}{3417} (\bibinfo{year}{1987}).

\bibitem[{\citenamefont{Lin and Falicov}(1966)}]{linandfalicov_66}
\bibinfo{author}{\bibfnamefont{P.~J.} \bibnamefont{Lin}} \bibnamefont{and}
  \bibinfo{author}{\bibfnamefont{L.~M.} \bibnamefont{Falicov}},
  \bibinfo{journal}{Phys. Rev.} \textbf{\bibinfo{volume}{142}},
  \bibinfo{pages}{441} (\bibinfo{year}{1966}).

\bibitem[{sil()}]{silas_08}
\emph{\bibinfo{title}{See {EPAPS} {D}ocument {N}o. [...] for the convergence
  properties of the {A}7~$\to$~sc phase transition. {F}or more information on
  {EPAPS}, see http://www.aip.org/pubservs/epaps.html.}}

\bibitem[{\citenamefont{Mujica et~al.}(2003)\citenamefont{Mujica, Rubio,
  Mu{\~{n}}oz, and Needs}}]{mujica_03}
\bibinfo{author}{\bibfnamefont{A.}~\bibnamefont{Mujica}},
  \bibinfo{author}{\bibfnamefont{A.}~\bibnamefont{Rubio}},
  \bibinfo{author}{\bibfnamefont{A.}~\bibnamefont{Mu{\~{n}}oz}},
  \bibnamefont{and} \bibinfo{author}{\bibfnamefont{R.~J.} \bibnamefont{Needs}},
  \bibinfo{journal}{Rev. Mod. Phys.} \textbf{\bibinfo{volume}{75}},
  \bibinfo{pages}{863} (\bibinfo{year}{2003}).

\bibitem[{\citenamefont{Li}(2003)}]{atomeye}
\bibinfo{author}{\bibfnamefont{J.}~\bibnamefont{Li}},
  \bibinfo{journal}{Modelling Simul. Mater. Sci. Eng.}
  \textbf{\bibinfo{volume}{11}}, \bibinfo{pages}{173} (\bibinfo{year}{2003}).

\bibitem[{\citenamefont{Wyckoff}(1963)}]{wyckoff}
\bibinfo{author}{\bibfnamefont{R.~W.~G.} \bibnamefont{Wyckoff}},
  \emph{\bibinfo{title}{Crystal Structures}}, vol.~\bibinfo{volume}{1}
  (\bibinfo{publisher}{John Wiley \& Sons, Inc.}, \bibinfo{address}{New York},
  \bibinfo{year}{1963}), \bibinfo{edition}{2nd} ed.

\bibitem[{\citenamefont{Falicov and Golin}(1965)}]{falicovandgolin_65}
\bibinfo{author}{\bibfnamefont{L.~M.} \bibnamefont{Falicov}} \bibnamefont{and}
  \bibinfo{author}{\bibfnamefont{S.}~\bibnamefont{Golin}},
  \bibinfo{journal}{Phys. Rev.} \textbf{\bibinfo{volume}{137}},
  \bibinfo{pages}{A871} (\bibinfo{year}{1965}).

\bibitem[{\citenamefont{Needs et~al.}(1986)\citenamefont{Needs, Martin, and
  Nielsen}}]{needs_86}
\bibinfo{author}{\bibfnamefont{R.~J.} \bibnamefont{Needs}},
  \bibinfo{author}{\bibfnamefont{R.~M.} \bibnamefont{Martin}},
  \bibnamefont{and} \bibinfo{author}{\bibfnamefont{O.~H.}
  \bibnamefont{Nielsen}}, \bibinfo{journal}{Phys. Rev. B}
  \textbf{\bibinfo{volume}{33}}, \bibinfo{pages}{3778} (\bibinfo{year}{1986}).

\bibitem[{\citenamefont{Mattheiss et~al.}(1986)\citenamefont{Mattheiss, Hamann,
  and Weber}}]{mattheiss_86}
\bibinfo{author}{\bibfnamefont{L.~F.} \bibnamefont{Mattheiss}},
  \bibinfo{author}{\bibfnamefont{D.~R.} \bibnamefont{Hamann}},
  \bibnamefont{and} \bibinfo{author}{\bibfnamefont{W.}~\bibnamefont{Weber}},
  \bibinfo{journal}{Phys. Rev. B} \textbf{\bibinfo{volume}{34}},
  \bibinfo{pages}{2190} (\bibinfo{year}{1986}).

\bibitem[{\citenamefont{Chang and Cohen}(1986)}]{chang_86}
\bibinfo{author}{\bibfnamefont{K.~J.} \bibnamefont{Chang}} \bibnamefont{and}
  \bibinfo{author}{\bibfnamefont{M.~L.} \bibnamefont{Cohen}},
  \bibinfo{journal}{Phys. Rev. B} \textbf{\bibinfo{volume}{33}},
  \bibinfo{pages}{7371} (\bibinfo{year}{1986}).

\bibitem[{\citenamefont{Littlewood}(1980)}]{littlewood_80}
\bibinfo{author}{\bibfnamefont{P.~B.} \bibnamefont{Littlewood}},
  \bibinfo{journal}{J. Phys. C} \textbf{\bibinfo{volume}{13}},
  \bibinfo{pages}{4855} (\bibinfo{year}{1980}).

\bibitem[{\citenamefont{Schirber and Van~Dyke}(1971)}]{schirber_71}
\bibinfo{author}{\bibfnamefont{J.~E.} \bibnamefont{Schirber}} \bibnamefont{and}
  \bibinfo{author}{\bibfnamefont{J.~P.} \bibnamefont{Van~Dyke}},
  \bibinfo{journal}{Phys. Rev. Lett.} \textbf{\bibinfo{volume}{26}},
  \bibinfo{pages}{246} (\bibinfo{year}{1971}).

\bibitem[{\citenamefont{Shang et~al.}(2007)\citenamefont{Shang, Wang, Zhang,
  and Liu}}]{shang_07}
\bibinfo{author}{\bibfnamefont{S.}~\bibnamefont{Shang}},
  \bibinfo{author}{\bibfnamefont{Y.}~\bibnamefont{Wang}},
  \bibinfo{author}{\bibfnamefont{H.}~\bibnamefont{Zhang}}, \bibnamefont{and}
  \bibinfo{author}{\bibfnamefont{Z.-K.} \bibnamefont{Liu}},
  \bibinfo{journal}{Phys. Rev. B} \textbf{\bibinfo{volume}{76}},
  \bibinfo{pages}{052301} (\bibinfo{year}{2007}).

\bibitem[{\citenamefont{Kawamura and Wittig}(1985)}]{kawamura_85}
\bibinfo{author}{\bibfnamefont{H.}~\bibnamefont{Kawamura}} \bibnamefont{and}
  \bibinfo{author}{\bibfnamefont{J.}~\bibnamefont{Wittig}},
  \bibinfo{journal}{Physica B} \textbf{\bibinfo{volume}{135}},
  \bibinfo{pages}{239} (\bibinfo{year}{1985}).

\bibitem[{\citenamefont{Chen et~al.}(1992)\citenamefont{Chen, Lewis, Su, Yu,
  and Cohen}}]{chen_92}
\bibinfo{author}{\bibfnamefont{A.~L.} \bibnamefont{Chen}},
  \bibinfo{author}{\bibfnamefont{S.~P.} \bibnamefont{Lewis}},
  \bibinfo{author}{\bibfnamefont{Z.}~\bibnamefont{Su}},
  \bibinfo{author}{\bibfnamefont{P.~Y.} \bibnamefont{Yu}}, \bibnamefont{and}
  \bibinfo{author}{\bibfnamefont{M.~L.} \bibnamefont{Cohen}},
  \bibinfo{journal}{Phys. Rev. B} \textbf{\bibinfo{volume}{46}},
  \bibinfo{pages}{5523} (\bibinfo{year}{1992}).

\bibitem[{\citenamefont{Da~{S}ilva and Wentzcovitch}(1997)}]{dasilva_97}
\bibinfo{author}{\bibfnamefont{C.~R.~S.} \bibnamefont{Da~{S}ilva}}
  \bibnamefont{and} \bibinfo{author}{\bibfnamefont{R.~M.}
  \bibnamefont{Wentzcovitch}}, \bibinfo{journal}{Comput. Mater. Sci.}
  \textbf{\bibinfo{volume}{8}}, \bibinfo{pages}{219} (\bibinfo{year}{1997}).

\bibitem[{\citenamefont{H{\"{a}}ussermann
  et~al.}(2002)\citenamefont{H{\"{a}}ussermann, S{\"{o}}derberg, and
  Norrestam}}]{haussermann_02}
\bibinfo{author}{\bibfnamefont{U.}~\bibnamefont{H{\"{a}}ussermann}},
  \bibinfo{author}{\bibfnamefont{K.}~\bibnamefont{S{\"{o}}derberg}},
  \bibnamefont{and}
  \bibinfo{author}{\bibfnamefont{R.}~\bibnamefont{Norrestam}},
  \bibinfo{journal}{J. Am. Chem. Soc.} \textbf{\bibinfo{volume}{124}},
  \bibinfo{pages}{15359} (\bibinfo{year}{2002}).

\bibitem[{\citenamefont{Durandurdu}(2005)}]{durandurdu_05}
\bibinfo{author}{\bibfnamefont{M.}~\bibnamefont{Durandurdu}},
  \bibinfo{journal}{Phys. Rev. B} \textbf{\bibinfo{volume}{72}},
  \bibinfo{pages}{073208} (\bibinfo{year}{2005}).

\bibitem[{\citenamefont{Feng et~al.}(2007)\citenamefont{Feng, Cui, Hu, and
  Liu}}]{feng_07}
\bibinfo{author}{\bibfnamefont{W.}~\bibnamefont{Feng}},
  \bibinfo{author}{\bibfnamefont{S.}~\bibnamefont{Cui}},
  \bibinfo{author}{\bibfnamefont{H.}~\bibnamefont{Hu}}, \bibnamefont{and}
  \bibinfo{author}{\bibfnamefont{H.}~\bibnamefont{Liu}},
  \bibinfo{journal}{Physica B} \textbf{\bibinfo{volume}{400}},
  \bibinfo{pages}{22} (\bibinfo{year}{2007}).

\bibitem[{\citenamefont{Zijlstra et~al.}(2008)\citenamefont{Zijlstra,
  Huntemann, and Garcia}}]{zijlstra_08}
\bibinfo{author}{\bibfnamefont{E.~S.} \bibnamefont{Zijlstra}},
  \bibinfo{author}{\bibfnamefont{N.}~\bibnamefont{Huntemann}},
  \bibnamefont{and} \bibinfo{author}{\bibfnamefont{M.~E.}
  \bibnamefont{Garcia}}, \bibinfo{journal}{New J. Phys.}
  \textbf{\bibinfo{volume}{10}}, \bibinfo{pages}{033010}
  (\bibinfo{year}{2008}).

\bibitem[{\citenamefont{Greene et~al.}(1995)\citenamefont{Greene, Luo, and
  Ruoff}}]{greene_95}
\bibinfo{author}{\bibfnamefont{R.~G.} \bibnamefont{Greene}},
  \bibinfo{author}{\bibfnamefont{H.}~\bibnamefont{Luo}}, \bibnamefont{and}
  \bibinfo{author}{\bibfnamefont{A.~L.} \bibnamefont{Ruoff}},
  \bibinfo{journal}{Phys. Rev. B} \textbf{\bibinfo{volume}{51}},
  \bibinfo{pages}{597} (\bibinfo{year}{1995}).

\bibitem[{\citenamefont{Iwasaki}(1997)}]{iwasaki_commentongreene_97}
\bibinfo{author}{\bibfnamefont{H.}~\bibnamefont{Iwasaki}},
  \bibinfo{journal}{Phys. Rev. B} \textbf{\bibinfo{volume}{55}},
  \bibinfo{pages}{14645} (\bibinfo{year}{1997}).

\bibitem[{\citenamefont{Iwasaki and Kikegawa}(1997)}]{iwasaki_97}
\bibinfo{author}{\bibfnamefont{H.}~\bibnamefont{Iwasaki}} \bibnamefont{and}
  \bibinfo{author}{\bibfnamefont{T.}~\bibnamefont{Kikegawa}},
  \bibinfo{journal}{Acta Crystallogr. Sect. B} \textbf{\bibinfo{volume}{53}},
  \bibinfo{pages}{353} (\bibinfo{year}{1997}).

\bibitem[{\citenamefont{Mc{M}ahon et~al.}(2000)\citenamefont{Mc{M}ahon,
  Degtyareva, and Nelmes}}]{mcmahon_00}
\bibinfo{author}{\bibfnamefont{M.~I.} \bibnamefont{Mc{M}ahon}},
  \bibinfo{author}{\bibfnamefont{O.}~\bibnamefont{Degtyareva}},
  \bibnamefont{and} \bibinfo{author}{\bibfnamefont{R.~J.}
  \bibnamefont{Nelmes}}, \bibinfo{journal}{Phys. Rev. Lett.}
  \textbf{\bibinfo{volume}{85}}, \bibinfo{pages}{4896} (\bibinfo{year}{2000}).

\bibitem[{\citenamefont{Degtyareva et~al.}(2004)\citenamefont{Degtyareva,
  Mc{M}ahon, and Nelmes}}]{degtyareva_04}
\bibinfo{author}{\bibfnamefont{O.}~\bibnamefont{Degtyareva}},
  \bibinfo{author}{\bibfnamefont{M.~I.} \bibnamefont{Mc{M}ahon}},
  \bibnamefont{and} \bibinfo{author}{\bibfnamefont{R.~J.}
  \bibnamefont{Nelmes}}, \bibinfo{journal}{Phys. Rev. B}
  \textbf{\bibinfo{volume}{70}}, \bibinfo{pages}{184119}
  (\bibinfo{year}{2004}).

\bibitem[{\citenamefont{Clark et~al.}(2005)\citenamefont{Clark, Segall,
  Pickard, Hasnip, Probert, Refson, and Payne}}]{castep}
\bibinfo{author}{\bibfnamefont{S.~J.} \bibnamefont{Clark}},
  \bibinfo{author}{\bibfnamefont{M.~D.} \bibnamefont{Segall}},
  \bibinfo{author}{\bibfnamefont{C.~J.} \bibnamefont{Pickard}},
  \bibinfo{author}{\bibfnamefont{P.~J.} \bibnamefont{Hasnip}},
  \bibinfo{author}{\bibfnamefont{M.~J.} \bibnamefont{Probert}},
  \bibinfo{author}{\bibfnamefont{K.}~\bibnamefont{Refson}}, \bibnamefont{and}
  \bibinfo{author}{\bibfnamefont{M.~C.} \bibnamefont{Payne}},
  \bibinfo{journal}{Z. Kristallogr.} \textbf{\bibinfo{volume}{220}},
  \bibinfo{pages}{567} (\bibinfo{year}{2005}).

\bibitem[{\citenamefont{Ceperley and Alder}(1980)}]{ceperley_80}
\bibinfo{author}{\bibfnamefont{D.~M.} \bibnamefont{Ceperley}} \bibnamefont{and}
  \bibinfo{author}{\bibfnamefont{B.~J.} \bibnamefont{Alder}},
  \bibinfo{journal}{Phys. Rev. Lett.} \textbf{\bibinfo{volume}{45}},
  \bibinfo{pages}{566} (\bibinfo{year}{1980}).

\bibitem[{\citenamefont{Perdew and Zunger}(1981)}]{perdew_81}
\bibinfo{author}{\bibfnamefont{J.~P.} \bibnamefont{Perdew}} \bibnamefont{and}
  \bibinfo{author}{\bibfnamefont{A.}~\bibnamefont{Zunger}},
  \bibinfo{journal}{Phys. Rev. B} \textbf{\bibinfo{volume}{23}},
  \bibinfo{pages}{5048} (\bibinfo{year}{1981}).

\bibitem[{\citenamefont{Perdew et~al.}(1996)\citenamefont{Perdew, Burke, and
  Ernzerhof}}]{perdew_96}
\bibinfo{author}{\bibfnamefont{J.~P.} \bibnamefont{Perdew}},
  \bibinfo{author}{\bibfnamefont{K.}~\bibnamefont{Burke}}, \bibnamefont{and}
  \bibinfo{author}{\bibfnamefont{M.}~\bibnamefont{Ernzerhof}},
  \bibinfo{journal}{Phys. Rev. Lett.} \textbf{\bibinfo{volume}{77}},
  \bibinfo{pages}{3865} (\bibinfo{year}{1996}).

\bibitem[{\citenamefont{Perdew and Wang}(1992)}]{perdew_92}
\bibinfo{author}{\bibfnamefont{J.~P.} \bibnamefont{Perdew}} \bibnamefont{and}
  \bibinfo{author}{\bibfnamefont{Y.}~\bibnamefont{Wang}},
  \bibinfo{journal}{Phys. Rev. B} \textbf{\bibinfo{volume}{45}},
  \bibinfo{pages}{13244} (\bibinfo{year}{1992}).

\bibitem[{\citenamefont{Vanderbilt}(1990)}]{vanderbilt_90}
\bibinfo{author}{\bibfnamefont{D.}~\bibnamefont{Vanderbilt}},
  \bibinfo{journal}{Phys. Rev. B} \textbf{\bibinfo{volume}{41}},
  \bibinfo{pages}{7892} (\bibinfo{year}{1990}).

\bibitem[{\citenamefont{Schiferl and Barrett}(1969)}]{schiferl_69}
\bibinfo{author}{\bibfnamefont{D.}~\bibnamefont{Schiferl}} \bibnamefont{and}
  \bibinfo{author}{\bibfnamefont{C.~S.} \bibnamefont{Barrett}},
  \bibinfo{journal}{J. Appl. Crystallogr.} \textbf{\bibinfo{volume}{2}},
  \bibinfo{pages}{30} (\bibinfo{year}{1969}).

\bibitem[{\citenamefont{Mehl}(2000)}]{mehl_00}
\bibinfo{author}{\bibfnamefont{M.~J.} \bibnamefont{Mehl}},
  \bibinfo{journal}{Phys. Rev. B} \textbf{\bibinfo{volume}{61}},
  \bibinfo{pages}{1654} (\bibinfo{year}{2000}).

\bibitem[{\citenamefont{Marzari et~al.}(1999)\citenamefont{Marzari, Vanderbilt,
  De~{V}ita, and Payne}}]{marzari_99}
\bibinfo{author}{\bibfnamefont{N.}~\bibnamefont{Marzari}},
  \bibinfo{author}{\bibfnamefont{D.}~\bibnamefont{Vanderbilt}},
  \bibinfo{author}{\bibfnamefont{A.}~\bibnamefont{De~{V}ita}},
  \bibnamefont{and} \bibinfo{author}{\bibfnamefont{M.~C.} \bibnamefont{Payne}},
  \bibinfo{journal}{Phys. Rev. Lett.} \textbf{\bibinfo{volume}{82}},
  \bibinfo{pages}{3296} (\bibinfo{year}{1999}).

\bibitem[{\citenamefont{Seifert et~al.}(1995)\citenamefont{Seifert, Hafner,
  Furthm{\"{u}}ller, and Kresse}}]{seifert_95}
\bibinfo{author}{\bibfnamefont{K.}~\bibnamefont{Seifert}},
  \bibinfo{author}{\bibfnamefont{J.}~\bibnamefont{Hafner}},
  \bibinfo{author}{\bibfnamefont{J.}~\bibnamefont{Furthm{\"{u}}ller}},
  \bibnamefont{and} \bibinfo{author}{\bibfnamefont{G.}~\bibnamefont{Kresse}},
  \bibinfo{journal}{J. Phys.: Condens. Matter} \textbf{\bibinfo{volume}{7}},
  \bibinfo{pages}{3683} (\bibinfo{year}{1995}).

\bibitem[{\citenamefont{Pickard and Needs}(2006)}]{pickard_06}
\bibinfo{author}{\bibfnamefont{C.~J.} \bibnamefont{Pickard}} \bibnamefont{and}
  \bibinfo{author}{\bibfnamefont{R.~J.} \bibnamefont{Needs}},
  \bibinfo{journal}{Phys. Rev. Lett.} \textbf{\bibinfo{volume}{97}},
  \bibinfo{pages}{045504} (\bibinfo{year}{2006}).

\bibitem[{\citenamefont{Birch}(1947)}]{birch_47}
\bibinfo{author}{\bibfnamefont{F.}~\bibnamefont{Birch}},
  \bibinfo{journal}{Phys. Rev.} \textbf{\bibinfo{volume}{71}},
  \bibinfo{pages}{809} (\bibinfo{year}{1947}).

\bibitem[{\citenamefont{Vinet et~al.}(1987)\citenamefont{Vinet, Smith,
  Ferrante, and Rose}}]{vinet_87}
\bibinfo{author}{\bibfnamefont{P.}~\bibnamefont{Vinet}},
  \bibinfo{author}{\bibfnamefont{J.~R.} \bibnamefont{Smith}},
  \bibinfo{author}{\bibfnamefont{J.}~\bibnamefont{Ferrante}}, \bibnamefont{and}
  \bibinfo{author}{\bibfnamefont{J.~H.} \bibnamefont{Rose}},
  \bibinfo{journal}{Phys. Rev. B} \textbf{\bibinfo{volume}{35}},
  \bibinfo{pages}{1945} (\bibinfo{year}{1987}).

\bibitem[{\citenamefont{Perdew and Wang}(1986)}]{perdewandwang_86}
\bibinfo{author}{\bibfnamefont{J.~P.} \bibnamefont{Perdew}} \bibnamefont{and}
  \bibinfo{author}{\bibfnamefont{Y.}~\bibnamefont{Wang}},
  \bibinfo{journal}{Phys. Rev. B} \textbf{\bibinfo{volume}{33}},
  \bibinfo{pages}{8800} (\bibinfo{year}{1986}).

\bibitem[{\citenamefont{Perdew}(1986)}]{perdew_86}
\bibinfo{author}{\bibfnamefont{J.~P.} \bibnamefont{Perdew}},
  \bibinfo{journal}{Phys. Rev. B} \textbf{\bibinfo{volume}{33}},
  \bibinfo{pages}{8822} (\bibinfo{year}{1986}).

\bibitem[{\citenamefont{Becke}(1988)}]{becke_88}
\bibinfo{author}{\bibfnamefont{A.~D.} \bibnamefont{Becke}},
  \bibinfo{journal}{Phys. Rev. A} \textbf{\bibinfo{volume}{38}},
  \bibinfo{pages}{3098} (\bibinfo{year}{1988}).

\bibitem[{\citenamefont{Needs et~al.}(1987)\citenamefont{Needs, Martin, and
  Nielsen}}]{needs_87}
\bibinfo{author}{\bibfnamefont{R.~J.} \bibnamefont{Needs}},
  \bibinfo{author}{\bibfnamefont{R.~M.} \bibnamefont{Martin}},
  \bibnamefont{and} \bibinfo{author}{\bibfnamefont{O.~H.}
  \bibnamefont{Nielsen}}, \bibinfo{journal}{Phys. Rev. B}
  \textbf{\bibinfo{volume}{35}}, \bibinfo{pages}{9851} (\bibinfo{year}{1987}).

\bibitem[{\citenamefont{Morosin and Schirber}(1972)}]{morosin_72}
\bibinfo{author}{\bibfnamefont{B.}~\bibnamefont{Morosin}} \bibnamefont{and}
  \bibinfo{author}{\bibfnamefont{J.~E.} \bibnamefont{Schirber}},
  \bibinfo{journal}{Solid State Commun.} \textbf{\bibinfo{volume}{10}},
  \bibinfo{pages}{249} (\bibinfo{year}{1972}).

\end{thebibliography}

\end{document}